\documentclass[preprint2]{aastex}
\usepackage{hyperref,xcolor,natbib}

\def\ltsim{\raisebox{-.4ex}{$\stackrel{<}{\sim}$}}

\def\twocolheads#1{\multicolumn{2}{c}{#1}}

\hypersetup{
    pdfnewwindow=true,      % links in new window
    colorlinks=true,        % false: boxed links; true: colored links
    linkcolor=blue,         % color of internal links
    citecolor=black,        % color of links to bibliography
    filecolor=blue,         % color of file links
    urlcolor=blue           % color of external links
}

\begin{document}

\title{James Webb Space Telescope can Detect Kilonovae in Gravitational Wave Follow-up Search}

\author{I. Bartos$^1$, T.L. Huard$^2$ and S. M\'arka$^1$}
\affil{$^1$Department of Physics, Columbia University, New York, NY 10027, USA}
\affil{$^2$Department of Astronomy, University of Maryland, College Park, MD 20742, USA}
\email{ibartos@phys.columbia.edu}

\begin{abstract}

Kilonovae represent an important electromagnetic counterpart for compact binary mergers, which could become the most commonly detected gravitational wave (GW) source. Follow-up observations, triggered by GW events, of kilonovae are nevertheless difficult due to poor localization by GW detectors and due to their faint near-infrared peak emission that has limited observational capability. We show that the Near-Infrared Camera (NIRCam) on the James Webb Space Telescope (JWST) will be able to detect kilonovae within the relevant GW-detection range of $\sim$\,200\,Mpc in short ($\ltsim$\,12-second) exposure times for a week following the merger.  Despite this sensitivity, a kilonova search fully covering a fiducial localized area of $10\,\mbox{deg}^2$ will not be viable with NIRCam due to its limited field of view. However, targeted surveys may be developed to optimize the likelihood of discovering kilonovae efficiently within limited observing time.  We estimate that a survey of $10\,\mbox{deg}^2$ focused on galaxies within 200\,Mpc would require about 13 hours, dominated by overhead times; a survey further focused on galaxies exhibiting high star-formation rates would require $\sim$5 hours. The characteristic time may be reduced to as little as $\sim$4 hours, without compromising the likelihood of detecting kilonovae, by surveying sky areas associated with 50\%, rather than 90\%, confidence regions of 3 GW events, rather than a single event. On detection and identification of a kilonova, a limited number of NIRCam follow-up observations could constrain the properties of matter ejected by the binary and the equation of state of dense nuclear matter.
\end{abstract}


\section{Introduction}

Compact binary mergers represent one of the most actively studied astrophysical phenomena. The mergers of neutron stars and stellar-mass black holes are promising targets for gravitational-wave (GW) observations \citep{2010CQGra..27q3001A}. They can also drive relativistic outflows, giving rise to some of the highest-energy emission of gamma rays, neutrinos and possibly cosmic rays.  The study of binary mergers can help us to further understand high-energy physics, the properties of matter at nuclear densities, stellar evolution, the abundance of heavy elements in the universe, as well as galaxy formation.

With the upcoming completion of advanced GW detectors, compact binary mergers are expected to be observed via GWs; such mergers will likely be the first sources from which GWs are directly detected. The construction of Advanced LIGO \citep{2010CQGra..27h4006H} will finish in 2015, with the detectors gradually reaching their design sensitivity by 2019 \citep{2013arXiv1304.0670L}. Advanced Virgo \citep{aVirgo} has a similar schedule with about a year delay. On reaching its design sensitivity, the Advanced LIGO-Virgo network could be able to detect a binary neutron star merger out to $\sim$\,450\,Mpc under favorable conditions, or, on average, out to $\sim$\,200\,Mpc \citep{2013CQGra..30l3001B,2013arXiv1304.0670L}. The expected detection rate of binary neutron star mergers with Advanced LIGO-Virgo's design sensitivity is $0.2-200$ per year \citep{2013arXiv1304.0670L}. Additionally, neutron star-black hole mergers will be detectable out to about a factor of two farther, corresponding to comparable detection rates \citep{2010CQGra..27q3001A}. The construction of additional GW observatories, such as KAGRA \citep{2012CQGra..29l4007S} and LIGO-India \citep{LIGOindia}, will further increase the number of detectable sources.

To maximize the sensitivity and scientific return of GW observation campaigns, GW candidates will be followed up by other instruments to find electromagnetic or neutrino counterparts. There are significant theoretical (e.g., \citealt{2012ApJ...746...48M,2011Natur.478...82N,2010CQGra..27q3001A}) and observational (e.g., \citealt{2008CQGra..25r4034K,2012A&A...541A.155A,2013APh....45...56S,2014MNRAS.443..738B,2014ApJS..211....7A}) efforts to identify promising counterparts.

Kilonovae are produced during the merger of a binary neutron star or a neutron star-black hole system \citep{1998ApJ...507L..59L,2005astro.ph.10256K,2005ApJ...634.1202R,2009aaxo.conf..312K,2010MNRAS.406.2650M}. If the neutron star is disrupted during the merger, the ejected neutron-rich matter produces heavy r-process elements, the radioactive decay of which powers the emission. Due to the large opacity at optical wavelengths of the formed r-process elements, kilonovae peak in the near-infrared with expected luminosities of $\sim10^{40}-10^{41}$\,erg\,s$^{-1}$, and last for over a week \citealt{2013ApJ...775...18B}. Nevertheless, this standard theoretical estimate has important uncertainties that may affect detectability. There is a significant ongoing theoretical effort to obtain a more complete understanding of the emission process, which may inform observations in the future (e.g., \citealt{2015MNRAS.450.1777K,2015MNRAS.446..750F}).

Kilonovae are promising electromagnetic counterparts of binary mergers because
(i) the emission is isotropic; therefore, the number of observable mergers is not limited by beaming; (ii) the week-long emission period allows sufficient time for follow-up observations; and (iii) once identified, source location can be accurately recovered, allowing for the identification of the host environment and the search for counterparts in other electromagnetic regimes. For comparison, some other electromagnetic counterparts, such as gamma-ray bursts and X-ray afterglows, are highly beamed, reducing the number of observable mergers. Optical and radio afterglow is also emitted off-axis; however, detection at large viewing angles is difficult \citep{2011ApJ...733L..37V}. Gamma-ray and X-ray emission are also short duration, which can also be a significant limitation as it allows less time for follow-up observations. On the other hand, the observability of kilonovae is currently limited by the lack of sufficiently sensitive survey instruments in the near-infrared band that can provide coverage over tens of square degrees, the typical area within which GW events will be localized by the Advanced LIGO-Virgo network \citep{2013arXiv1304.0670L}. The sole kilonova observation so far took advantage of the precisely reconstructed source direction \citep{2013Natur.500..547T,2013ApJ...774L..23B}.

The James Webb Space Telescope (JWST) is a planned, highly sensitive infrared space telescope with an expected launch in 2018. In this paper, we show that its Near Infrared Camera (NIRCam; \citealt{2004SPIE.5487..628H}), with a spectral range of 0.6-5\,$\mu$m, is well suited for quickly detecting kilonovae following GW triggers from Advanced LIGO-Virgo.

In Section \ref{section:sensitivity}, we determine the sensitivity of NIRCam to detecting kilonova emission within its field of view. In Section \ref{section:motivation}, we motivate the need for observation strategies that target nearby galaxies to host kilonovae; in Section \ref{section:target}, we discuss examples of such targeted surveys to identify kilonovae.  The possible role of other astrophysical sources being misidentified as kilonovae are presented in Section \ref{section:identification}.  In Section \ref{section:sciencereach}, we investigate how NIRCam could probe the kilonova emission model and source parameters. We summarize our findings in Section \ref{section:summary}.

%\section{JWST/NIRCam sensitivity to kilonovae}
\section{NIRCam sensitivity to kilonovae}
\label{section:sensitivity}

We first calculate the JWST/NIRCam sensitivity in detecting kilonovae. We determine which NIRCam filter is the most sensitive for the expected emission spectrum, and derive the integration time necessary for a 10\,$\sigma$ detection. Only then may we develop viable observing strategies with the greatest likelihood of enabling kilonovae identification.

We adopt the kilonova emission model obtained by \cite{2013ApJ...775...18B} using time dependent, multi-wavelength radiative transport calculations. \cite{2013ApJ...775...18B} simulated a range of emission parameters, in particular ejecta masses $\sim 10^{-3}-10^{-1}$M$_{\odot}$ and characteristic ejection velocities $\beta\equiv v/c \sim 0.1-0.3$. These ranges seem to cover the expected and
%%%%%the
observed kilonova emission parameters \citep{2013Natur.500..547T,2013ApJ...774L..23B}.

To determine the observed kilonova flux at Earth, we assume a source luminosity distance of $200$\,Mpc, which represents the average reach of the Advanced LIGO-Virgo network at design sensitivity to binary neutron star mergers. While some sources may occur even closer or as far as $\sim$\,450\,Mpc under the most favorable conditions, our choice of 200\,Mpc provides a good estimate of the potential limitations of a follow-up search with NIRCam.

With kilonova spectra expected to peak in the near-infrared, we calculated the minimum integration time sufficient to detect a kilonova for each of NIRCam's wide-band filters. Preliminary filter sensitivity curves were taken from JWST's webpage\footnote{\url{http://www.stsci.edu/jwst/instruments/nircam/instrumentdesign/filters}}. The two wide-band filters with the longest wavelengths [F356W, F444W] were excluded from the analysis because the simulations of \cite{2013ApJ...775...18B} did not fully cover these bands, leaving us with filters F070W, F090W, F115W, F150W, F200W and F277W\footnote{For each filter, the number in the filter name represents 100 times the central wavelength in microns.  For example, the central wavelength of F356W is $\sim$\,3.56 $\mu$m.)}. Nevertheless, as we will see, F200W and F277W are already sufficiently sensitive, and F277W is less sensitive than F200W soon after the binary merger, indicating that F356W and F444W are likely not more optimal for kilonova detection.

Emulating the JWST NIRCam Exposure Time Calculator\footnote{\url{http://jwstetc.stsci.edu/etc/input/nircam/imaging/}}, we determined the minimum time needed for the different filters to obtain a 10\,$\sigma$ detection as a function of the time after a binary merger at 200\,Mpc for the range of kilonova emission parameters. In particular, for each filter, we first calculated the signal-to-noise
(SNR) of a kilonova as a function of the (i) total exposure time and
(ii) time-varying source flux, determined by integrating the kilonova flux
density weighed with the filter transmittance. We then selected that exposure time associated with 10$\sigma$, and repeated the process for a range of elapsed time since merger.

\begin{figure}[t!]
\begin{center}
\resizebox{0.47\textwidth}{!}{\includegraphics[trim=0in 0.3in 0in 0.6in]{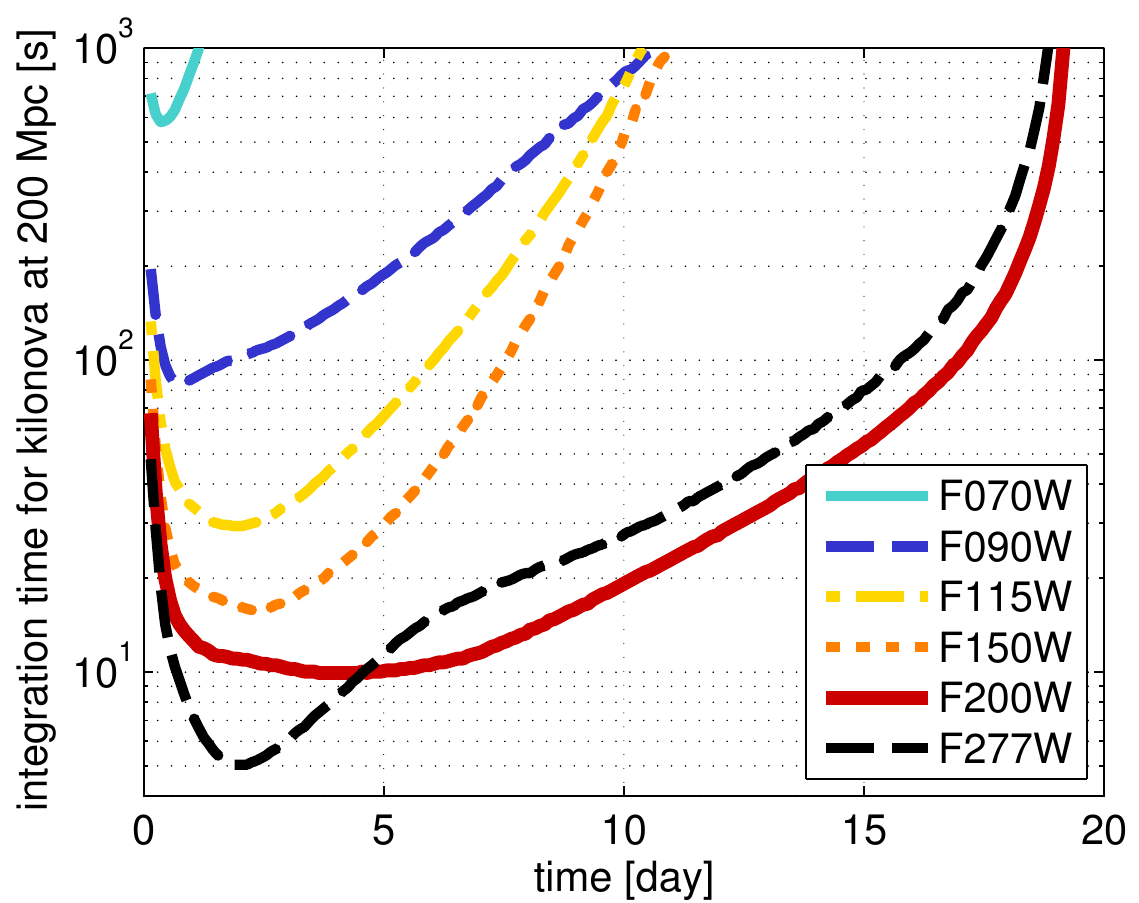}}
\end{center}
\caption{Integration time for JWST NIRCam necessary to detect a kilonova ($10^{-2}$\,M$_\odot$ ejected mass, $\beta=0.2$) with 10\,$\sigma$ significance as a function of time after a binary merger at 200\,Mpc, for different NIRCam filters. The results indicate that the F200W and F277W filters are well suited to observe kilonovae, depending on how quickly the observations are obtained following the merger.}
\label{JWSTmindetectionduration}
\end{figure}

Results for a kilonova with $10^{-2}$\,M$_\odot$ ejected mass and $\beta=0.2$ characteristic velocity are shown in Fig. \ref{JWSTmindetectionduration} as an illustration of our findings, in general.  One can see that integration times of only (i) 5--10 seconds are necessary in F277W if the kilonova is seen within 5 days; and, (ii) 10--12 seconds are necessary in F200W if it is caught within a week. JWST/NIRCam will have unprecedented sensitivity to detect kilonovae.

\section{Full survey of GW trigger events}
\label{section:motivation}

While NIRCam has unprecedented sensitivity, there are several limiting factors, relevant for a kilonova search aiming to cover a large, contiguous region of sky:
(i) NIRCam has a minimum integration time of 10.6~seconds due to the readout time of the detectors in full-array mode; and (ii) the 5\arcsec\ gaps between the four detectors of each module and the 50\arcsec\ gap between the two modules, which necessitate positional shifts of less than half the 2.2\arcmin\,$\times$\,4.4\arcmin\ field of view for full coverage in the short wavelength channel in survey mode.  These positional shifts would also serve to provide, for each point imaged, 2--3 dithered exposures, which may be used to remove image artifacts, such as bad pixels and cosmic rays.
To uniformly cover a fiducial localized sky area of $10$\,deg$^2$ associated with a GW source, at a confidence level of 90\%, these two factors would necessitate $\sim$\,50\,hours of total exposure time, \emph{not including overheads}, making it unfeasible to cover the full region associated with a GW trigger event.

\section{Targeted surveys}
\label{section:target}

\begin{table*}[t!]
\caption{Targeted Surveys of GW Trigger Event}
\label{surveytimes}
\begin{center}
\begin{tabular}{lcccc}
\hline
\hline  \\[-9pt]
							& \twocolheads{\underline{90\% confidence region}}	& \twocolheads{\underline{50\% confidence region}}	\\
							& All 			& H$\alpha$	& All 			& H$\alpha$	\\
							& galaxies	& galaxies	& galaxies	& galaxies	\\
\hline \\[-9pt]
Fiducial sky area [deg$^2$]					& 10		& 10		& 2		& 2	\\
Number of Galaxies							& 80		& 30		& 16		& 6	\\
Slew Time to First Galaxy [min]					& 30.0	& 30.0	& 30.0	& 30.0	\\
Slew Time to Each Subsequent Galaxy [min]		& 3.6		& 4.5		& 3.6		& 4.5	\\
Guide Star Acquisition Time for Each Galaxy [min]	& 4.0		& 4.0		& 4.0		& 4.0	\\
Dithering Time for Each Galaxy$^a$ [min]		& 1.0		& 1.0		& 1.0		& 1.0	\\
%\hline \\[-9pt]
Other Overhead Time$^b$ [min]				& 25.1	& 10.1	& 5.9		& 2.9\\
\hline \\[-9pt]
Total Overhead Time [hr]						& 12.3	& 5.3		& 8.5$^c$		& 4.3$^c$	\\
Total Exposure Time [min]					& 16.0	& 6.0		& 9.6$^c$		& 3.6$^c$\\
\hline \\[-9pt]
Total Time [hr]								& 12.6	& 5.4		& 8.7$^c$		& 4.4$^c$	\\
\hline  \\[-9pt]
\end{tabular}
\end{center}
\footnotesize{{$^a$ Two dither motions, each requiring 30 seconds, in order to obtain three dithered images for each galaxy.}}\\
\footnotesize{{$^b$ Includes 65 seconds for filter move and detector configuration, necessary only once, and an adopted detector deadtime of 18 seconds for each galaxy.  The detector deadtime for our subarray observations was assumed to be 6 seconds, based on discussion in \cite{gordon2012}, for each of the three images of a galaxy.}}\\
\footnotesize{{$^c$ Total times listed for targeted survey of 50\% confidence regions are those associated with {\it{three}} GW trigger events for direct comparison to total times for targeted survey of the 90\% confidence region of a single event, yielding a similar likelihood of kilonova detection.}}\\
\end{table*}

Several approaches may be used to develop a viable NIRCam survey, triggered by a GW source, to search for a kilonova most efficiently and with the greatest chance of success.  The survey should be triggered only by the most promising GW sources:  those with high signal-to-noise ratios and well localized areas. While we consider a fiducial sky area of $10\,$deg$^2$, a substantial fraction of GW trigger events will have larger reconstructed sky areas (e.g., \citealt{2013arXiv1304.0670L,2014ApJ...795..105S}) and therefore require proportionally longer NIRCam observing times. In this section, we discuss approaches that yield targeted surveys in order to reduce the required NIRCam observing time.

Instead of covering the full $10$\,deg$^2$ area, one approach is to focus the follow-up search toward known galaxies \citep{2008ApJ...675.1459K,2008CQGra..25r4034K,2013ApJ...767..124N,2014ApJ...784....8H,2041-8205-801-1-L1} within $\sim$\,200\,Mpc, taking advantage of the expectation that binary mergers occur in or near galaxies \citep{2010ApJ...708....9F}. Following \cite{2013ApJ...767..124N}, the number density of galaxies within 200\,Mpc is estimated to be $\sim$\,8\,deg$^{-2}$; therefore, one would need to follow up $\sim$\,80 galaxies for a fiducial GW sky area of $10$\,deg$^2$.

Since most galaxies can be covered well within a 400\,$\times$\,400\,pixel$^2$ (13\arcsec\,$\times$\,13\arcsec) region (e.g., \citealt{2010ApJ...708....9F}), NIRCam can be used in a \emph{subarray} imaging mode (e.g., \citealt{2014arXiv1411.1754B}) for such a survey of galaxies. In this mode, exposure times as short as $\sim$2 seconds are possible. In principle, this minimum exposure could be used with F277W for each of 3 dithered images, to mitigate image artifacts, of a galaxy if observed within 3 days of the merger.  For our purposes, we consider instead 3 dithered 4-second exposures, enabling the use of F200W for up to a week after the merger.  For the 80 galaxies, assuming the most time-intensive case of only one galaxy present in a field, the total exposure time would be 16 minutes, not including overheads.  While such an approach represents a significant improvement over full coverage of the GW event, the time associated with overheads is expected to be significant.

The overheads for such a NIRCam imaging survey may be attributed to a number of sources, summarized in Table~\ref{surveytimes} (\citealt{gordon2012}).  The most significant contributors to overhead times would be the time for JWST to slew and acquire guide stars. The standard assumption for slew time to a new field is 30 minutes, which we adopt for the slew to the first galaxy following a GW event.  Slew times to subsequent galaxies associated with that event depend on the angular separation between the galaxies.
While JWST slews at a nominal rate of at least 90$^\circ$ per hour, the slew rates are non-linear with distance due to inertia.  Shorter slew distances are associated with slower effective rates.  A survey of 80 galaxies over a 10 deg$^2$ region suggests an average separation of $\sim$\,21\arcmin\ between nearby galaxies.  According to expectations for slew rates as a function of slew distance (\citealt{Gardner2010WhitePaper}), our typical slew time would be no more than $\sim$\,3.6 minutes\footnote{We discuss this slew time as an upper limit since it represents an extrapolation from 2.5 minutes for a 5\arcmin\ slew and 3.3 minutes for a 17\arcmin\ slew (\citealt{Gardner2010WhitePaper}).  Since greater slews are more efficient, the actual slew time may be less than that extrapolated.  Note that the 2.5 minutes for the 5$\arcmin$ slew with the 4 minutes for the guide star acquisition is comparable to mission requirement MR-180 (\citealt{JWSTMissionReq}), reflected also in the operations requirement MO-442 (\citealt{JWSTOperationsReq}), that a 4.7$\arcmin$ slew, including guide star acquisition, be accomplished in 8 minutes or less.} to point from one galaxy to the next. This typical slew is sufficiently large that each field will require a different guide star and 4 minutes for its acquisition.  Accounting for the move to the first galaxy, moves to subsequent galaxies, and guide star acquisitions for all fields, the total overhead time associated with slewing and guide stars is 10.6 hours. The next most significant contributor to overhead times would be the time associated with the small (e.g., $<$\,2\arcsec) dithering, which totals 1.3 hours for all fields.  Finally, other overheads include moving the filter wheel, detector configuration and deadtime, which total only 0.4 hour. With the full overhead time of 12.3~hours, the total time for this targeted survey of 80 galaxies in $10$\,deg$^2$, associated with a GW trigger event, is 12.6~hours.

A second approach is to further focus the survey to target only those nearby galaxies most likely to host kilonovae.  For example, the star formation rate is correlated with the rate of compact binary mergers \citep{2010ApJ...725.1202L}; H$\alpha$ is an indicator of the star formation rate and therefore may be used to optimize the probability of discovering kilonovae efficiently within a limited observing time.  Specifically, H$\alpha$ mapping may be used to identify those galaxies responsible for 90\% of star formation and for 50\% of the mass (\citealt{2013ApJ...764..149M}; \citealt{2041-8205-801-1-L1}).  Based on the galaxy stellar mass function for the nearest galaxies (\citealt{2012MNRAS.421..621B}; \citealt{2013ApJ...767...50M}), with the component due to star-forming galaxies normalized such that they account for 60\% of mass of the nearest galaxies (\citealt{2013ApJ...764..149M}), these H$\alpha$-detected galaxies would represent about 40\% of the total number of galaxies.  Thus, with only $\sim$\,30 galaxies most likely to host kilonovae within 200\,Mpc over $10$\,deg$^2$, and typical separations of $\sim$\,35\,\arcmin, the total overhead associated with slewing and guide star acquisition is further decreased to no more than 4.7 hours.  With the additional overheads, the total time for this targeted survey would be 5.4 hours.

We note that a focus on H$\alpha$ serves to solve another problem:  a comprehensive galaxy catalog complete to 200\,Mpc may not be ready by 2019, when Advanced LIGO detectors will reach their design sensitivities.  For reference, recent galaxy catalogs are estimated to be only $\sim$\,60\% complete to 100\,Mpc (e.g., \citealt{2011CQGra..28h5016W}).  \cite{2041-8205-801-1-L1} suggested that on-the-fly H$\alpha$ mapping may be used to generate catalogs of galaxies for kilonova searches triggered by GW events, and found that mapping a $\sim$\,$10$\,deg$^2$ region to identify these galaxies within 200\,Mpc can be done cost effectively and quickly, within a day, using a 1-2 meter class telescope.

Finally, another approach for a targeted survey is to focus on the galaxies in the 50\% (as opposed to 90\%) confidence region of the localized sky area associated with a GW event.  Such an approach can decrease the covered sky area by a factor of $\sim$\,5 \citep{2014ApJ...795..105S}.  In this case, if the targeted survey follows up three GW triggers instead of one, then a similar probability of success (90\%) is achieved by covering 40\% fewer galaxies than the survey of a single 90\% confidence region.  Such an approach, however, does not require 40\% less time since the slew to the first galaxy following a GW trigger would occur three times.  Targeted surveys of all nearest ($\le$\,200\,Mpc) galaxies and only nearby H$\alpha$ galaxies within 50\% confidence regions of the localized areas of three optimal GW sources could be done with NIRCam within a total of 8.7~hours and 4.4~hours, respectively. Not only would such an approach require less time, but it would also result in fewer galaxies needing followup observations to confirm a kilonova identification.

JWST is not an efficient facility for a survey involving quick exposures of many fields.  In these examples of targeted surveys, the overhead times are 50--70 times greater than the exposure times. Despite this inefficiency, by judiciously choosing the galaxies most likely to host kilonovae associated with GW events, the NIRCam observing time can be minimized to enable high-impact science not possible from other facilities.

\section{Identification of kilonovae}
\label{section:identification}

Kilonovae and supernovae are among the intrinsically brightest extragalactic compact sources, more than two orders of magnitude brighter than other, more typical sources at optical wavelengths (e.g., \citealt{2013IAUS..281....9K}; \citealt{2009PASP..121.1334R}).  In the near-infrared, kilonovae are even brighter (\citealt{2013ApJ...775...18B}).  A single-epoch NIRCam observation is therefore sufficient to identify any galaxy associated with a kilonova candidate, particularly a candidate identified within a week of and coincident with a GW trigger event.

Once a kilonova candidate is identified, one needs to distinguish it from the foreground (e.g., Milky Way stars and asteroids), background (e.g., distant galaxies), and unresolved (e.g., HII regions) sources. For example, multiple background galaxies that are sufficiently bright may overlap a nearby ($\sim200$\,Mpc) galaxy, resulting in a continuous quasi-point source with brightness comparable to kilonovae. Distinguishing between a bona-fide kilonova and continuous sources requires a template be obtained, either before or after the occurrence of the kilonova, or multiple NIRCam exposures obtained days apart.  For the former, there is currently no suitable all-sky survey for this purpose, but it will be possible to carry out a survey selectively for the relevant galaxies, using other facilities to complement NIRCam observations.  For the latter, NIRCam followup observations could be obtained on the galaxies.  This approach may not be optimal since it requires double JWST time, but the advantage is that such observations would provide the template to distinguish kilonova from continuous sources and enable characterization of its dimming.

Some transient sources (e.g., foreground asteroids and dwarf novae, background supernovae) could be mistaken as kilonovae, resulting in false positive transient events.
To address this issue, we first recall the results of \cite{2009aaxo.conf..312K}, which finds that the background rate of supernovae, aligned by chance with a nearby galaxy, for Advanced LIGO-Virgo in a single snapshot within 12\,deg$^2$ with r-band luminosity of $r < 24$ should be $\lesssim 0.1$. That study also finds the foreground rate of flares to be even less, $\lesssim 0.01$. These sources are expected to be the dominant source of false positive events. While these false positive event rates could, in principle, be different in NIRCam's infrared wavelength range, they are unlikely to be significantly greater. We therefore expect that any survey focused on galaxies, and especially the targeted surveys discussed in \S\ref{section:target}, will render the number of false positive transient events practically negligible.

\section{Probing the kilonova emission model}
\label{section:sciencereach}

The detection of kilonovae from a compact binary merger detected via GWs will help answer a number of important questions. Many of these questions may be addressed with the detailed observation of one kilonova, making even the limited observation time required for one detection valuable. If one determines the peak flux and the timescale for the kilonova to reach
maximum light, one can deduce the quantity of r-process ejecta and its
mean velocity (e.g., \citealt{2010MNRAS.406.2650M}).  For a statistical sample of
such events, this information will address whether neutron star mergers
are the dominant source for producing r-process elements in our Galaxy.
When coupled with GW measurements of the parameters of the
merging binary, the mass and velocity of the ejecta will also help
constrain the equation of state of dense nuclear matter (e.g., \citealt{2013ApJ...773...78B}).

Determining the peak flux of a kilonova may nevertheless not be possible unless the light curve is observed multiple times during the emission period.  In principle, such rapid follow-up observations are possible.  The initial survey, executed as a target of opportunity, could occur within 2 days (operations requirement MO-210; \citealt{JWSTOperationsReq}) of the GW event.  With calibrated, processed NIRCam images then available within 5 days after downlink (e.g., operations requirement MO-41; \citealt{JWSTOperationsReq}), follow-up NIRCam observations of a candidate or confirmed kilonova could then be executed as a target of opportunity again within 2 days.  Thus, if an investigator develops data processing pipelines such that an appropriate gravitational wave event can be selected and NIRCam images be searched in relatively short periods of time, two epochs of NIRCam observations
separated by a week may be obtained for a kilonova within 9--10 days of the binary merger.  This time frame is likely sufficiently short to enable additional observations of the kilonova, providing information on its temporal evolution.

Triggered {\it
spectroscopic} follow-up, either with JWST itself or a large ground-based
near-infrared telescope (e.g., the Giant Magellan Telescope), could
confirm the merger origin of the event by detecting the absorption lines
of exotic r-process elements (e.g., \citealt{2014arXiv1411.3726K}).  The strength of
individual lines, once identified, could in principle be used to determine
the relative abundances of individual nuclei.

\section{Conclusion}
\label{section:summary}

We showed that JWST/NIRCam can easily detect kilonovae out to distances relevant to GW observations of compact binary mergers. To efficiently survey the sky for kilonovae following a GW detection, NIRCam observations will need to be directed toward galaxies within the GW distance range. For a maximum source distance of 200\,Mpc, which will be the typical distance of detected binaries with Advanced LIGO-Virgo at design sensitivity, the required NIRCam observation time for a fiducial $10$ deg$^2$ GW-localized sky area would be about 13 hours overwhelmingly dominated by slewing and guide star acquisition. This time can be significantly decreased, perhaps to about 4 hours, by focusing on (i) the galaxies with the highest star formation rates and (ii) especially those within the most probable GW sky area, enabling a survey of multiple GW detections in less time without compromising detection of a kilonova counterpart.

We find that the identification of kilonovae is unlikely to be limited by foreground or background transient events. With such detection capability, JWST/NIRCam will be able to (i) regularly survey GW event candidates for kilonova counterparts, therefore establishing a statistically significant sample size of kilonova emission parameters, and (ii) will be able to rapidly find kilonovae, enabling detailed study of their temporal and spectral evolution.

\vspace{4 mm}
The authors thank Brian Metzger, Alessandra Corsi, Jonah Kanner, and an anonymous referee for their helpful suggestions. They also thank Klaus Pontoppidan for discussions and clarifications of the overheads and operations associated with JWST/NIRCam imaging, and Tracy Beck for explanation of timing requirements associated with target-of-opportunity observations and data processing.  This paper was approved for publication by the LIGO Scientific Collaboration. IB and SM are thankful for the generous support of Columbia University in the City of New York and the National Science Foundation under cooperative agreement PHY-0847182.

\bibliographystyle{apj}
%\bibliography{Ref_JWST}

\end{document}